\documentclass{tRES2e}
\usepackage{subfigure}
\usepackage{color}
\begin{document}



\title{Adapting astronomical source detection software to help detect animals in thermal images obtained by unmanned aerial systems}

\author{S.~N.~Longmore$^{\rm a}$$^{\ast}$\thanks{$^\ast$Corresponding author. Email: s.n.longmore@ljmu.ac.uk
\vspace{6pt}}, R. P. Collins$^{\rm a}$, S. Pfeifer$^{\rm a}$, S. E. Fox$^{\rm a}$, M. Mulero-P\'{a}zm\'{a}ny$^{\rm b}$, F. Bezombes$^{\rm c}$, A. Goodwin$^{\rm d}$, M. de Juan Ovelar$^{\rm a}$, J.~H. Knapen$^{\rm e,f}$ and S.~A.~Wich$^{\rm g, h}$ \\\vspace{6pt}  $^{a}${\em{Astrophysics Research Institute, Liverpool John Moores University, UK}};\\
$^{b}${\em{Departamento de Ciencias Naturales, Universidad T\'{e}cnica Particular de Loja, San Cayetano Alto, Loja, Ecuador}};\\
$^{c}${\em{General Engineering Research Institute, Liverpool John Moores University, UK}}\\
$^{d}${\em{Remote Insight Basecamp, Creative Campus, Jamaica Street, Liverpool, UK}}\\
$^{e}${\em{Instituto de Astrof\'\i sica de Canarias, E-38200 La Laguna, Tenerife, Spain}}\\
$^{f}${\em{Departamento de Astrof\'\i sica, Universidad de La Laguna, E-38205 La Laguna, Tenerife, Spain}}\\
$^{g}${\em{School of Natural Sciences and Psychology, Liverpool John Moores University, UK}}\\
$^{h}${\em{Institute for Biodiversity and Ecosystem Dynamics, University of Amsterdam, Sciencepark 904, Amsterdam 1098, the Netherlands}};\\\received{v4.2 released February 2014} }

\maketitle

\begin{abstract}

In this paper we describe an unmanned aerial system equipped with a thermal-infrared camera and software pipeline that we have developed to monitor animal populations for conservation purposes. Taking a multi-disciplinary approach to tackle this problem, we use freely available astronomical source detection software and the associated expertise of astronomers, to efficiently and reliably detect humans and animals in aerial thermal-infrared footage. Combining this astronomical detection software with existing machine learning algorithms into a single, automated, end-to-end pipeline, we test the software using aerial video footage taken in a controlled, field-like environment. We demonstrate that the pipeline works reliably and describe how it can be used to estimate the completeness of different observational datasets to objects of a given type as a function of height, observing conditions etc. -- a crucial step in converting video footage to scientifically useful information such as the spatial distribution and density of different animal species. Finally, having demonstrated the potential utility of the system, we describe the steps we are taking to adapt the system for work in the field, in particular systematic monitoring of endangered species at National Parks around the world.


\end{abstract}

\section{Introduction}

A major task for the conservation research community is  monitoring of species' distribution and density. Species monitoring has usually been undertaken by surveys on the ground (either on foot or by car), from the air with manned aircraft, and more recently from space using satellites \citep{buckland2001,buckland2004, fretwell2012, mcmahon2014}. There is a wealth of data from ground and aerial surveys, and the analytical methods for analyses of such data have been well developed \citep{buckland2001,buckland2004}. However, the costs of these surveys are high due to the extensive time commitments involved and the small areas which can be covered by individual ground surveys. Although aerial surveys cover larger areas, the costs of hiring or purchasing aircraft are often cost prohibitive for conservation research and/or aircraft are simply not available in the areas that need to be surveyed. In addition, flying low over areas where landing opportunities are limited in case of an emergency is risky \citep{sasse2003}. Alternative methods are therefore urgently needed to monitor biodiversity better. 

A particularly promising method for biodiversity monitoring is the use of drones\footnote{We use the terms ``drone" and ``unmanned aerial vehicle" (UAV) to denote the vehicle and  ``unmanned aerial system" (UAS) to specify the vehicle plus imaging payload.}. The falling cost of both drones and small, hi-fidelity cameras which can be attached to these drones has led to an explosion in the use of aerial footage for conservation research. Many of these applications require detecting and identifying objects in the obtained images. The fact that this task is mostly conducted manually -- which is labour-intensive, inherently slow and costly -- is a major bottleneck in maximising the potential of the enormous volumes of data being collected, and the efficiency with which drones can be used. 
 
To date most drone work has been done with cameras operating at visible wavelengths \citep[e.g.,][]{jones2006, rodriguez2012, koh2012, barasona2014, linchant2015, wich2015, mulero2015, van2015, canal2016}. Studies at these wavelengths suffer from two limitations. Firstly, visible cameras only work in the day time, so are essentially ``blind" for half of the time. Certain applications, such as identifying poaching activity or tracking and monitoring the large number of species that are active at night, are therefore impossible. Secondly, because the light that we do see at visible wavelengths is reflected sunlight, all objects have very similar brightness. This makes it difficult and computationally expensive to separate objects from the surrounding background in an automated way without human intervention, adding to the challenge of efficiently detecting and identifying objects of interest in the data. 

Because the body temperature of most warm-blooded animals is approximately $\sim$300\,K, using detectors that are optimally sensitive to emission from objects at this temperature might improve detection of animals. The spectral energy distributions of objects with temperatures around 300\,K peak at wavelengths of $\sim$10\,$\mu$m. Cameras operating in this ``thermal"-infrared regime are therefore optimally sensitive to wavelengths at which warm-blooded animals emit most of their radiative energy. At these same wavelengths, cooler objects -- such as plants and surrounding terrain -- rarely emit strongly. In theory, the resulting large intensity contrast between warm-blooded animals and the background makes the thermal regime particularly well suited to easily finding and identifying warm-blooded animals, both during the day and at night. This advantage of thermal infrared imaging is maximised in cold, dry environments and diminishes as the temperature and humidity increase \citep[see e.g.,][]{mulero2014}. 

Although the thermal-infrared regime offers major advantages, significant barriers have hampered monitoring at these wavelengths. Until recently thermal-infrared cameras have been prohibitively expensive for small-scale research projects. Now that they are becoming affordable (a basic unit can be purchased at the $\pounds1-10$\,k level),  researchers in many different areas are exploiting the advantages they offer to detect and identify species such as red deer \citep{chretien2016}, American bison, fallow deer, gray wolves, and elks \citep{chretien2015}, koala \citep{gonzalez2016}, roe deer \citep{israel2011} and hippos \citep{lhoest2015}. \citet{mulero2014}  compare the efficacy of aerial footage at both thermal and visible wavelengths for rhinoceros surveillance and people detection in the frame of anti-poaching operations. 

As the volume of data from such studies increases, the major bottleneck in maximising the scientific output are the analysis tools, which are currently less well developed than at visible wavelengths. Indeed, many studies have used manual detection and identification of species, which can be prohibitively time consuming and the results are dependent on the person doing the detection and identification. This inherent subjectivity means the results may not be repeatable, which is clearly not ideal. Several different groups have been working to overcome this problem by developing algorithms to automate the detection and identification process \citep[e.g.,][]{lhoest2015, chretien2015, chretien2016, gonzalez2016}. 

In this paper we seek to use a complementary method to help overcome this bottleneck. Astronomers have routinely been using (thermal-)infrared images for a number of decades to derive the properties of astrophysical objects. Due to the extreme distances involved, the emission from astronomical objects is usually very faint, and objects are often not well resolved (i.e. objects are only marginally larger than pixel scale of the camera and point spread function of the telescope). The analytical tools developed by astronomers are therefore optimised to overcome these limitations -- exactly the same challenges that need to be overcome for monitoring animal populations in aerial video footage. 

Last year we began a new research venture aiming to build upon astronomical software algorithms, develop an automated thermal (infrared) drone detection/identification pipeline and demonstrate its potential for aiding conservation research by monitoring animal populations. In this paper we first summarise the system we constructed before describing initial results of a pilot proof-of-concept project to assess the system's applicability for field work. We finish by outlining future prospects for expanding the system to tackle challenges in conservation research. 

\section{Materials and methods}
\label{sec:system_description}

\begin{figure}
 \centering
 \includegraphics[width=\columnwidth]{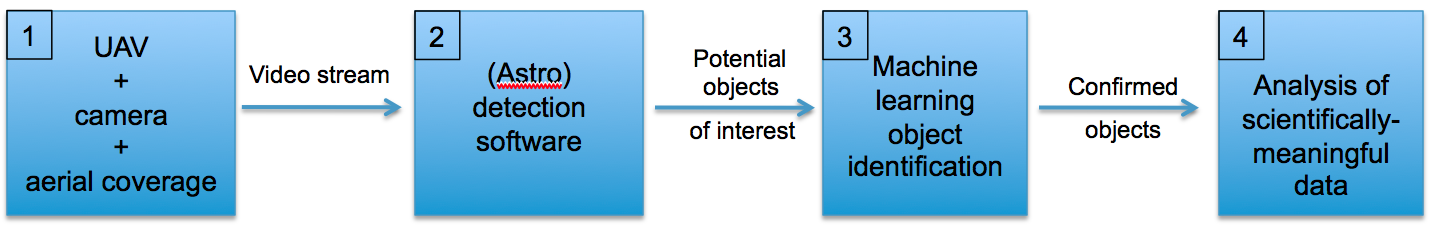}
 \caption{Flow chart of the system.}
 \label{fig:flow_chart}
\end{figure}

Figure~\ref{fig:flow_chart} shows a schematic flow chart of the four different components making up the system. In this section we briefly describe each of these components and outline the design decisions made when putting the system together. In $\S$\ref{sec:pilot} we provide more specific details for the system used in the proof-of-concept project.

Component 1 relates to the hardware (the unmanned aerial vehicle (UAV) and camera) and the flight path (height, trajectory, velocity) used to obtain the video footage. As some aspect of Component 1 will vary every time the system is used, and the subsequent detection/identification software components are sensitive to variations in the input video data, it is important to optimise the pipeline for the specific hardware and flight plan. Details of the hardware used for the pilot project, and the way in which details of the hardware plus flight plan can be used to optimise the pipeline are described in $\S$\ref{sec:pilot}.

Components 2 and 3 are the main software parts of the system that were developed. It was the lack of existing freely available software pipelines to automatically detect and identify objects in (thermal) video data that originally motivated the project. After investigating potential solutions to developing the end-to-end pipeline, it became clear that there are a plethora of freely available software packages that solve individual parts of the problem. Given the long-term goal of the project to develop a robust and cost-effective system, when presented with different software choices, we opted for freely available, open source, widely used and tested software as a top priority. A secondary consideration was the run time and resources required to achieve a given task, opting for the quicker and less intensive option as this would be preferable if attempting a real-time pipeline in the future (see $\S$~\ref{sec:future}). After investigating different solutions we opted for Python as the language of choice for its portability, large number of pertinent open source packages that could be of potential use for the pipeline, and ease of combining the different components into a single, coherent pipeline. In $\S$~\ref{sub:detection_software} and \ref{sub:identification_software} below, we describe Components 2 and 3 in detail.

\subsection{Source detection software}
\label{sub:detection_software}

As outlined in the introduction, astronomers have been analysing (thermal-) infrared images for decades and there are many different software packages available to aid such analysis. Given the criteria outlined in $\S$~\ref{sec:system_description}, for the source detection software of this project (Component 2) we used the routines available in {\sc astropy}, the standard freely available, open source astrophysical software package for Python that has been widely-used and tested within the astrophysics community \citep{astropy}. Specifically, we used routines within {\sc photutils} to identify and extract sources within images. 

There are several different functions available: {\sc irafstarfind}, {\sc daofind} and {\sc find\_peaks}. Comparing these functions, {\sc irafstarfind} and {\sc daofind} deliver a larger range of customisation of target source parameters and return more detail on detected sources. However, for this particular application these details were deemed unnecessary and also increased the run time significantly. We found that the {\sc find\_peaks} routine provides the best compromise in terms of functionality and run time.

 {\sc find\_peaks} works by identifying sub-regions in a 2D image where all the pixels within a user-defined area of $N$ pixels are above a user-specified threshold value. The optimal values for these parameters depend on many factors, most critically the absolute level and variation in background ``noise" across the image, and the size, shape and separation of the objects of interest. In $\S$~\ref{sec:pilot} we describe how the pilot project was used to assess how well these key parameters can be determined on-the-fly and how robust the pipeline detections are against variations in the initial estimate of these parameters.
 
\subsection{Source identification software}
\label{sub:identification_software}

There are several freely available source identification methods written in Python. The most prominent and well-supported revolve around the {\sc OpenCV}  (Open Source Computer Vision) libraries. {\sc OpenCV} contains many packages, most of which proved unsuitable for our purposes. In particular, we found (perhaps unsurprisingly) that the libraries dedicated to human facial recognition provided poor results when attempting to identify humans and animals in thermal aerial footage. In addition, these libraries took the longest to train. This left one obvious candidate, the {\sc hog} (histogram of oriented gradients) detector, which uses Support Vector Machine (SVM) and has had great success in the field of human detection since its inception \citep{dalal2005}.

The computer vision libraries rely on machine learning algorithms. In order to correctly identify objects, one must first ``train'' these algorithms. This is done by providing the libraries with two different sets of images: one containing images of the object of interest and the other containing images in a similar setting but without containing the object of interest. The libraries then process these different lists of images to calculate a set of ``vectors" which optimally describe the object of interest and that can be used to identify the object in subsequent images. The more fully the original set of images covers the possible range of viewing angles, distances etc. to the object of interest, the more robust the identification process will be. 

We conducted some initial tests running the machine learning code directly on the full image sizes from different cameras. The hope was that if this could be run quickly and accurately enough, it could alleviate the need for the detection step (Component 2, $\S$~\ref{sub:detection_software}). However, it was immediately obvious that when most of the pixels in most of the video frames do not contain objects of interest (as will usually be the case), this is an incredibly inefficient way to operate. The run time of the machine learning step is orders of magnitude longer than the detection step, so even on a high-end desktop machine the run time is prohibitive. This motivated our decision to stick with an initial detection step, and then use cutouts of potential detections as the input for the machine learning step, thereby only running the computationally most expensive algorithms on a very small subset of the full dataset.

In $\S$~\ref{sec:pilot} we discuss the machine learning and training process in the context of the pilot study.

\section{Results}
\label{sec:pilot}

When working in real monitoring situations, video footage will likely cover a range of environments such as varying topography and vegetation coverage. However, for the pilot  stage of the project our goal was to obtain footage that could be most readily used to test and develop the software pipeline. We therefore opted to obtain footage under the simplest possible environmental conditions -- relatively uniform and limited vegetation coverage with little topographic variation -- and containing only a single, easily identifiable species of animal. We worked with a local farmer who gave us permission to fly a drone over the cows in his fields and take such test data.

The data for the pilot project were taken on the afternoon of 14 July 2015 at Arrowe Brook Farm Wirral, UK (53.3701 N, -3.1105). The weather was mostly sunny with a few clouds. The average air temperature during the flights was $\sim15-20^\circ$C. The rational for flying in the UK during the day in summer is that this provides similar ground temperatures to the night time ground temperatures at many of the world's major national parks involved in conservation of megafauna.

\begin{figure}
 \centering
 \includegraphics[width=\columnwidth]{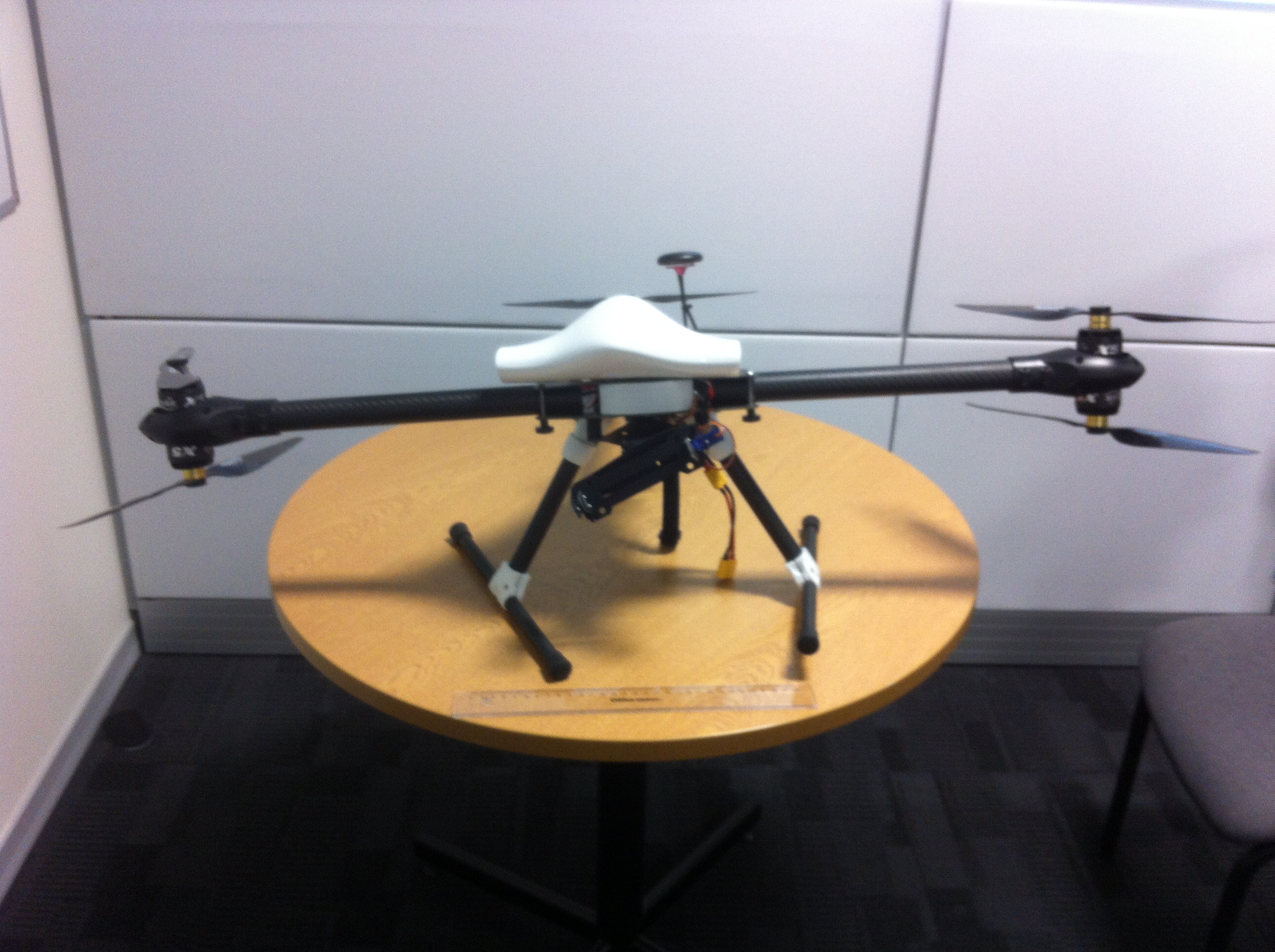}\\
  \includegraphics[width=\columnwidth]{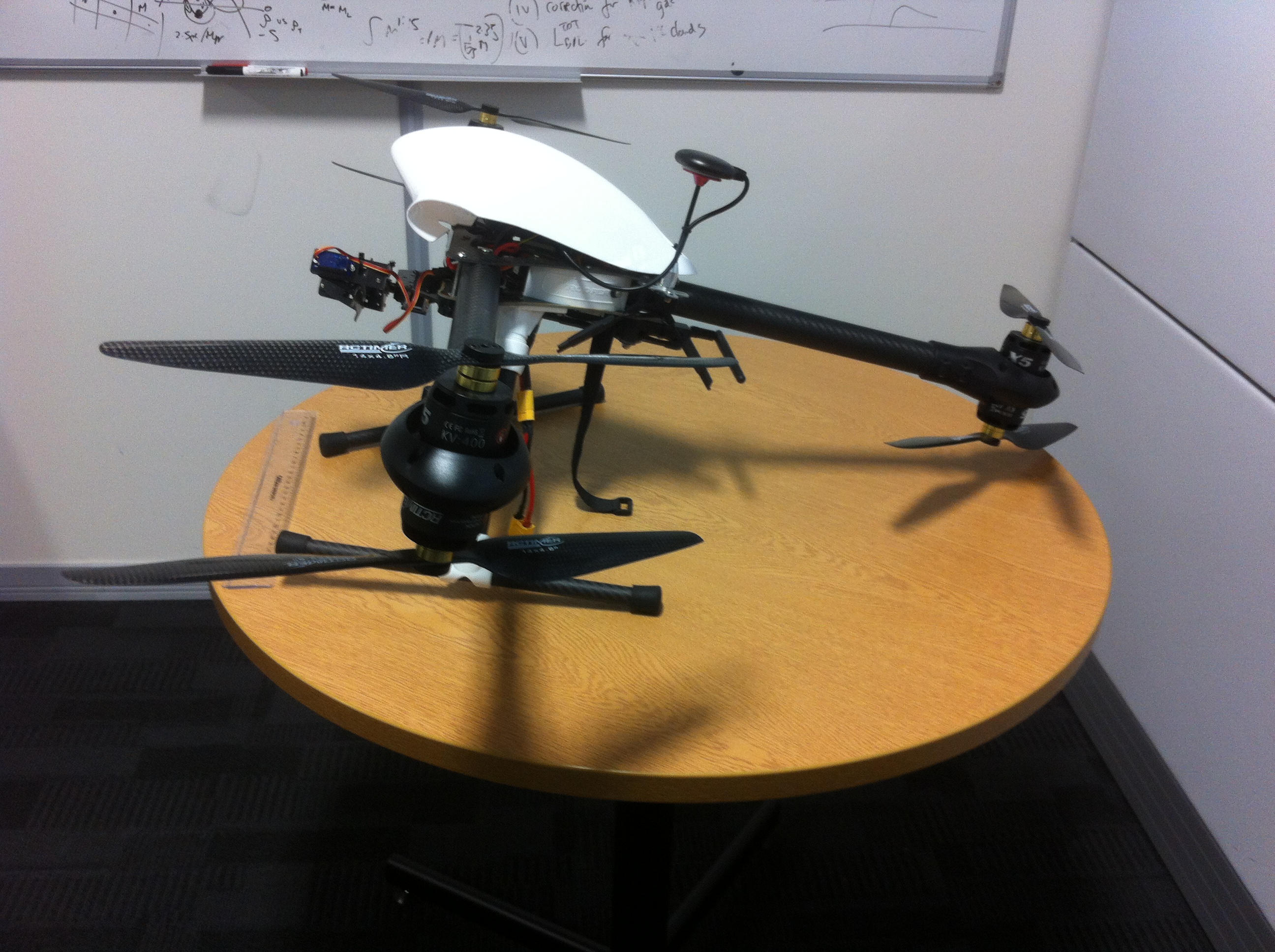}
 \caption{Photographs of the drone used in the pilot project. A standard 30-cm, clear plastic ruler is shown on the table to provide the scale.}
 \label{fig:drone_photo}
\end{figure}

The UAV for the pilot project was a custom-made 750\,mm carbon-folding Y6-multi-rotor with an APM 2 autopilot from 3Drobotics shown in Figure~\ref{fig:drone_photo}. The system uses 14" x 4.8" carbon fibre propellors and X5 3515 400\,Kv motors. A FrSky Taranis transmitter and receiver were used. The system was powered with a Multistar 6S 8000\,mAh lipo battery. The 3DR radio telemetry kit was used at 433\,MHz. As ground station we used a Lenovo laptop. The system was provided by http://www.droneshop.biz. The camera was a passively cooled, FLIR, Tau 2 LWIR Thermal Imaging Camera Core with a 7.5mm lens and 640$\times$512 pixels operating at 75\,Hz. For this first test run, the camera was strapped to a gimbal at the front of the UAV, put into video mode, and switched on manually to begin taking footage before the UAV took off. The video footage was captured directly on a camera's detachable universal serial bus (USB) data storage system (ThermalCapture) that was developed by http://www.teax-tec.de/. We conducted two flights with this system to capture both humans and cows from different angles and heights, up to a maximum altitude of 120\,m. After the flights were finished, the data were then manually transferred to a separate computer for reduction and analysis.

\subsection{Optimising the detection algorithm}
\label{sub:optimise_detection_pilot}

The next step was to run the {\sc find\_peaks} source detection software on the video footage. It became immediately clear that blindly running {\sc find\_peaks} on the data with no constraints on the input parameters produced very poor results. We found that the two key parameters to set are the threshold noise level above which pixels are considered as potential detections, and the minimum/maximum area allowed for an object to be selected as a potential detection. As we describe below, with some basic knowledge about (i) the characteristic properties of the sources of interest; (ii) the camera angle with respect to the horizon, (iii) the camera field of view and pixel scale, and (iv) the height of the UAS above ground level as a function of time, it is possible to provide sensible inputs to the {\sc find\_peaks} detection algorithm that produce good detection statistics.

\begin{figure}
 \centering
 \includegraphics[width=\columnwidth]{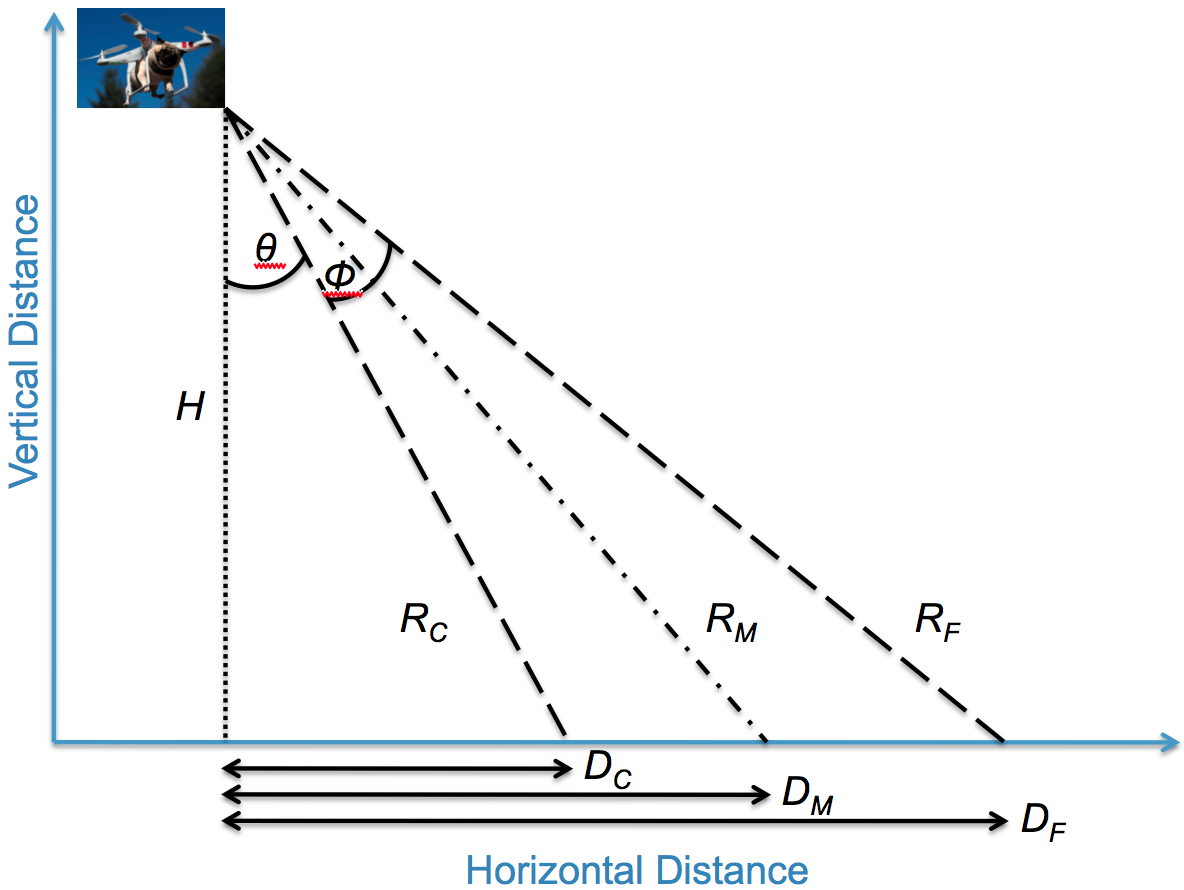}
 \caption{Schematic diagram showing distances and geometric projections. See $\S$~\ref{sub:optimise_detection_pilot} for definition of the variables.}
 \label{fig:drone_fig_distances}
\end{figure}

Figures~\ref{fig:drone_fig_distances} shows a simplified schematic diagram of the drone plus camera system geometry with respect to the ground through which one can determine required inputs\footnote{See \citet{mulero2014} for a similar geometric relations approach for analysing aerial video footage including equations app.}. In this schematic, the ground is approximated as being flat. At any given moment the drone is at an instantaneous height, $H$, above the ground. $H$ is known accurately as a function of time from global positioning system (GPS) data recorded on both the camera and the UAV. The camera is at an angle $\theta$ with respect to the ground, where $\theta = 0^\circ$ corresponds to pointing the camera straight at the ground, and $\theta = 90^\circ$ corresponds to the camera pointing straight ahead towards the horizon. This angle is either set and recorded by the gimbal if using a moving camera setup, or fixed for the flight and can therefore be measured prior to launch. The angle $\phi$ is the camera's field of view which is known to high accuracy from the design specifications. 

\begin{figure}
 \centering
 \includegraphics[width=\columnwidth]{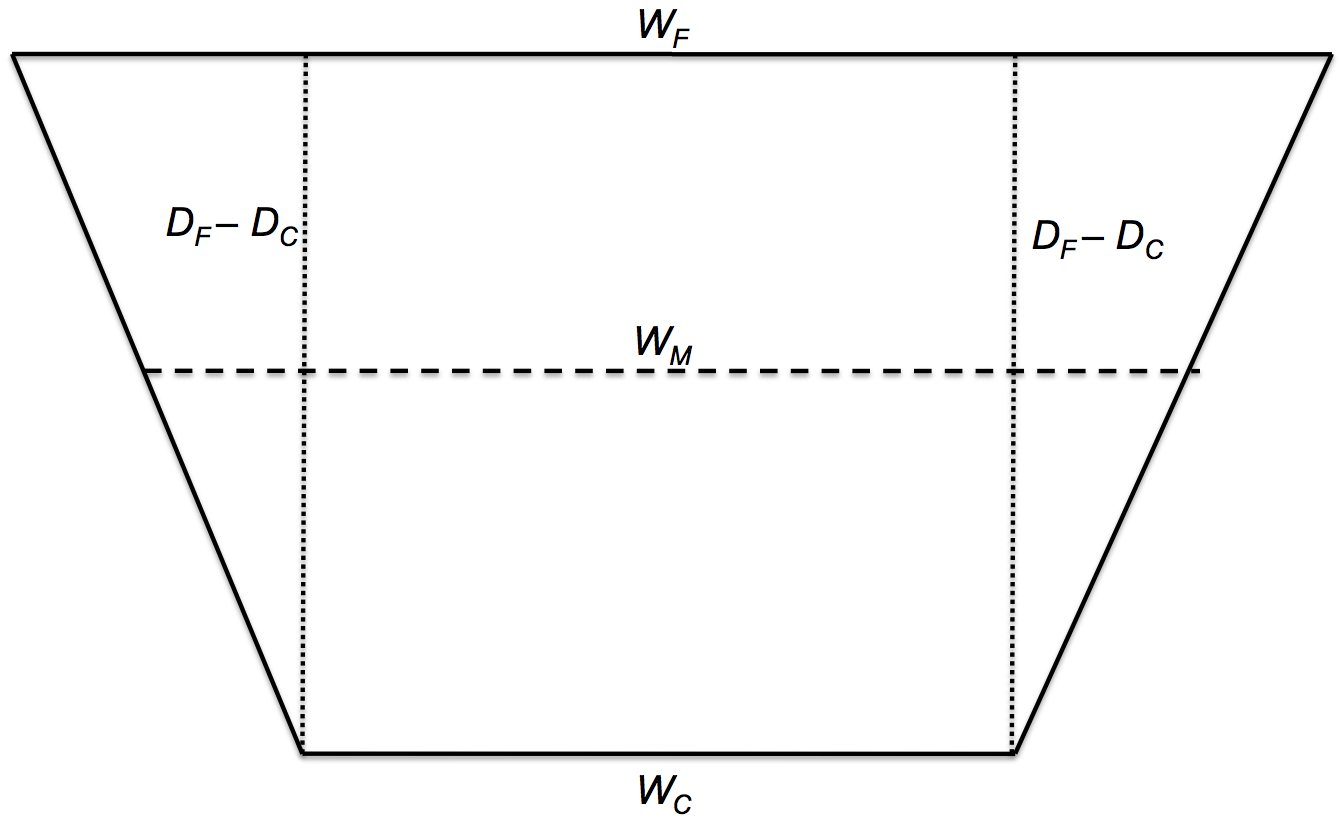}
 \caption{Schematic diagram showing the camera's projected field of view on the ground. See $\S$~\ref{sub:optimise_detection_pilot} for definition of the variables.}
 \label{fig:drone_fig_field_of_view}
\end{figure}

We next consider the instantaneous area of ground viewed by the camera attached to the drone. The camera frame is rectangular with 640$\times$512 pixels. Each of the pixels subtends an equal angular extent. Geometric projection means the area of ground covered is not rectangular, but rather trapezoidoidal in shape, as shown in Figure~\ref{fig:drone_fig_field_of_view}.

With $H$, $\theta$, $\phi$ and the pixel scale known, simple trigonometric relations provide the distance from the camera and the horizontal distance along the ground -- which we denote $R$ and $D$, respectively, in Figure~\ref{fig:drone_fig_distances} --  to any point on the ground in the camera's field of view. The subscripts $C$, $M$, $F$ refer to the closest, middle and farthest distances from the camera, respectively. 

We can now use a rough estimate of the expected range of size and shape of the objects of interest to constrain the angular area of these objects that will be projected into the camera's field of view. For example, in the pilot project we were interested in detecting cows and humans in the video footage. We estimated the plausible maximum and minimum height, length and width of cows and humans vary between 0.5\,m and 2\,m. The angular size of an individual object projected on to the camera will depend both on the intrinsic size of the object and the angle from which it is viewed. We calculated the maximum and minimum projected area for cows and humans, and used these as the range in input areas for the detection algorithm. Note that this area varies as a function of position across the camera, decreasing for objects closer to the horizon. We found that using this simple estimation of expected angular extent of objects did a good job of only selecting objects of interest in the data. 

Clearly there will always be some uncertainties in $H$, $\theta$, $\phi$ and the pixel scale, and the assumption of flat ground will break down in certain situations. We  attempted to simulate the effect of these uncertainties by over/under estimating the expected projected area of objects by varying amounts. We found that the number of sources detected was robustly recovered even when over/under estimating the source size by a factor of two. This is much larger than the expected over/under estimate due to  uncertainties in $H$, $\theta$, $\phi$ and the pixel scale, and should be robust in all but the most extreme topographic environments (e.g., cliffs, mountains).

With knowledge of the ground temperature and expected animal temperature, and assuming their spectral energy distributions are well described as a blackbody emitting radiation at that temperature, one can use the Planck function to calculate the radiative intensity of both animals and ground. This in turn provides the expected contrast,  and hence approximate intensity level above the background noise to set as the threshold for qualifying a detection.  Making the standard assumption that the flux from an object falls off as the reciprocal of the distance squared, we can also estimate how the expected flux may change as a function of the location on the camera. We found that although this approach worked, it was in fact easier just to calculate a simple root mean square (RMS) deviation in the intensity level of all the pixels in the image and take a value $3-5$ times higher as the threshold. Both approaches did a good job of finding the objects of interest in the data and the robustness of the detection did not depend strongly on the exact threshold level, as long as it was far enough above the background noise level.

In summary, with reasonable estimates of the source size and intensity the detection algorithm did a reliable job of recovering the objects of interest in the data. Comfortingly, the number of detections is robust when the size and intensity are over-/under-estimated by up to factors of 2. However, it should be noted that the pilot experiment was conducted under idealised conditions, and the conclusions need to be reassessed in real field conditions.

\subsection{Optimising the identification algorithm}
\label{sub:optimise_identification_pilot}

We then wanted to test the robustness of the machine learning object identification code and understand the conditions under which it worked efficiently. In order to do this we first ran all the individual frames of the video footage from one of the flights through the detection process described in $\S$~\ref{sub:optimise_detection_pilot}. At the location of every detection we cut out a small, fixed size\footnote{The size of the cutout was determined to be twice the maximum extent in pixels of the objects as determined in $\S$~\ref{sub:optimise_detection_pilot}}, square subset of the full frame around that detection, and stored all of these cutout images in a folder of `potential identifications'. We then inspected each of these by eye and sorted which of these contained humans, cows or neither (i.e., spurious detections). This provided the ground truth results against which we could test the machine learning algorithms. 

We then selected video frames in which there were no cows or humans, and randomly extracted cutouts with the same size to act as the negative images for the training step of the machine learning process. Using the human and cow cutouts as the `positive' training images and the random background noise images as the `negative' training images, we ran the {\sc hog$+$svm} algorithm to generate human and cow vectors that could be used to identify cows and humans in other video footage.

We then attempted to run Components 2 and 3 of the pipeline concurrently on video footage from other, different flights. Using the steps outlined in $\S$~\ref{sub:optimise_detection_pilot} we generated a set of cutout images of all detections, which were classed as `potential identifications'.  We then ran these potential identifications through the machine learning code, separated them into those which the code thought did or did not contain cows and humans, and subsequently verified by eye whether the code had identified these correctly. In this way were able to quantify the accuracy of detections in each frame.

We found that the variable that most strongly affected the accuracy was the height of the drone above the ground. When the drone was close to the ground ($\lesssim$80\,m altitude in the particular example of the pilot study data), the cows and humans were well resolved and the algorithm did a reasonable job of correctly identifying them. The average detection accuracy at low altitudes was at least $\sim$70\% with a scatter of roughly $\pm 10$\% depending on variation in background levels. The majority of cases in which the algorithm mis-classified an actual detection as a non-detection was when the cows were huddled close together, and there were multiple overlapping cows. These issues were exacerbated when the drone altitude increased, and at a height of 120m the average successful identification fraction had dropped by 10\%.

Clearly, dealing with such issues is an area for improvement for further versions of the pipeline. However, it was encouraging to note that while in an individual frame the algorithm may only correctly identify 70\% of the cows, over the course of the full video footage, all of the cows were individually detected in a large number of frames. With appropriate tracking of individually identified objects as a function of time, it should be possible to achieve much higher accuracy for the number of animals correctly identified.  

\begin{figure}
 \centering
 \includegraphics[width=\columnwidth]{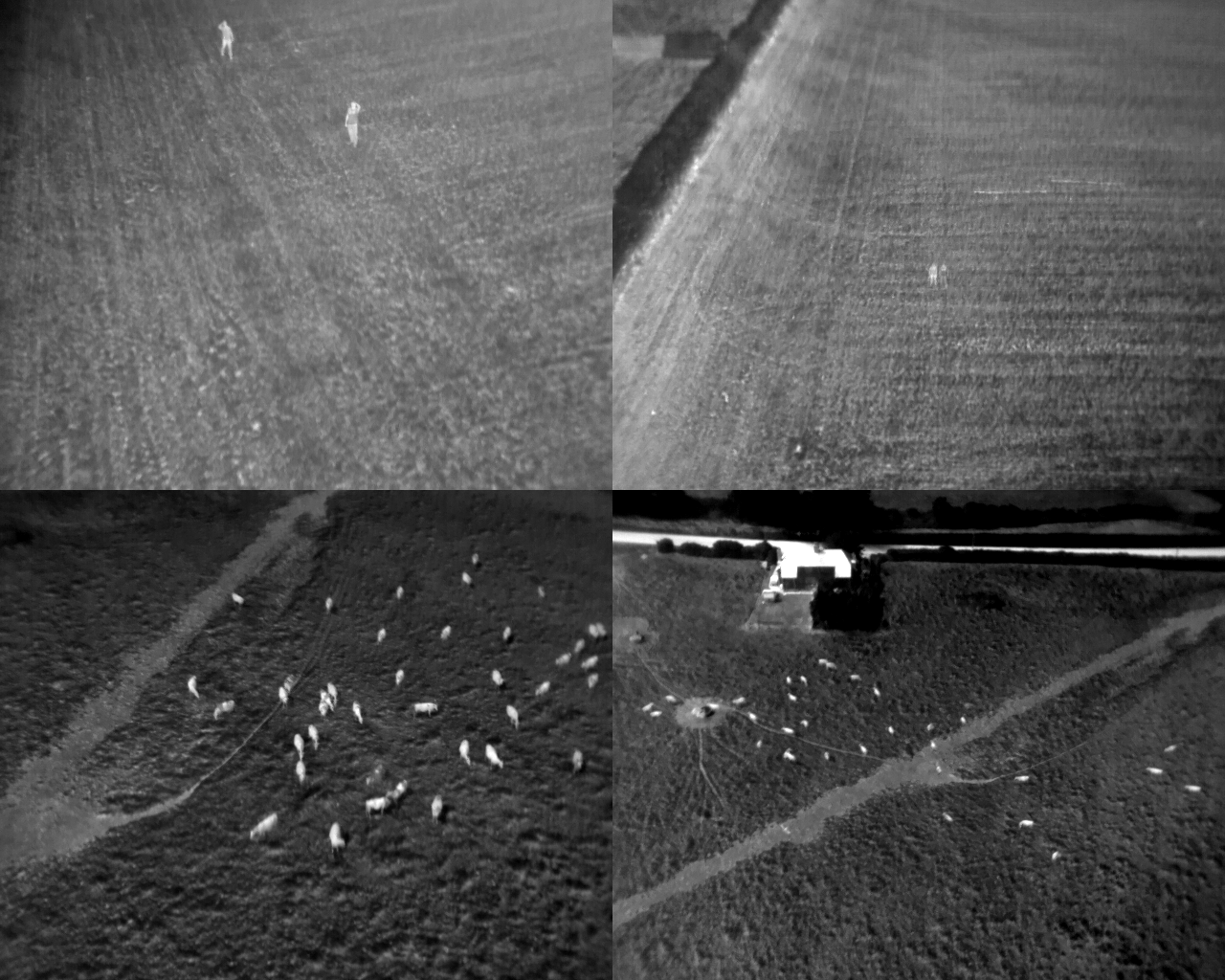}
 \caption{Single frame thermal-infrared snapshots from the pilot project of humans (top row) and cows (bottom row). The left and right hand columns shows low and high altitude footage, respectively. }
 \label{fig:thermal_snapshots}
\end{figure}

\begin{figure}
 \centering
 \includegraphics[width=\columnwidth]{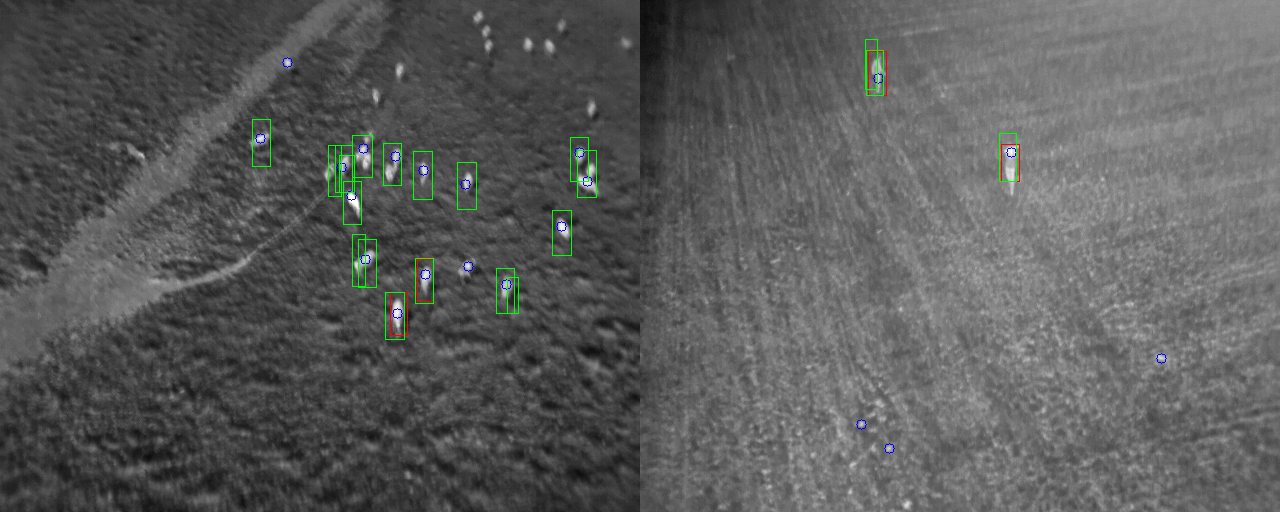}
 \caption{Example output from the detection and identification pipeline towards cows (left) and humans (right). Blue circles denote `hot-spots' that were initially selected by the detection algorithm before the size criteria was applied. Green and red rectangles show the location of objects which the algorithm ultimately identified as either  a cow (left) or human (right). For illustrative purposes, in this instance the geometric correction for the expected size as a function of pixel location has not been applied. This shows that the cows in the background (i.e. towards the top of the image) with a smaller projected size were not automatically identified.}
 \label{fig:cows_and_human_detections}
\end{figure}

\section{Discussion}
\label{sec:optimisation}

\subsection{Optimising the scientific return from future monitoring programs}
\label{sub:optimisation}

From the pilot project we have begun to quantify under what circumstances our approach will be able to robustly detect and identify objects of interest. This in turn can be used to help guide decisions regarding the appropriateness of using this system in different situations, and how to optimise the scientific return from monitoring programs.

For example, in any aerial monitoring program, there will always be a tradeoff in the area of ground that can be covered in a single flight and the level of detail with which it is possible to survey that area. Put simply, flying closer to the ground will provide higher resolution images at the expense of the area that can be covered. The key to optimising the scientific return lies in calculating the maximum height at which to fly before it is no longer possible to detect or identify the objects of interest robustly.

Using the pilot project as an example, if we wanted to monitor cows over a much larger area with the same system as above, we could use the information gained from the pilot project to determine how much area it will be feasible to cover. We have determined that the optimal drone height for detection and identification of cows is 80\,m. From Figure~\ref{fig:drone_fig_field_of_view}, at a height of 80\,m the camera's field of view corresponds to an area on the ground of approximately 17500\,m$^2$. Assuming that the drone can fly for 30 minutes (the approximate lifetime of the battery) and flies at an average speed of 15\,ms$^{-1}$ the drone can cover a ground area of order 3.5\,km$^2$ per flight. This allows one to optimise the flight paths for the region to be monitored.

Flying height will, however, not only be determined by resolution requirements, but also by the potential reaction of animals to flying a UAS. The cows did not show any visible reaction to the UAS flying above them, but other studies have indicated that in some cases animals can have increased heart rates  or show flight responses when UASs are flown close to them \citep{ditmer2015, rummler2015}. In a review on wildlife conservation and UASs, \citet{chabot2015} conclude that generally UASs lead to no or low levels of disturbance, and specifically so when compared to direct surveys on the ground or much more noisy surveys from manned aircraft. At the moment there are no general ethical guidelines for using UASs in animal surveys, but conservationist have proposed the development of such guidelines \citep{hodgson2016}. The ever-increasing resolution of satellites and opportunities for monitoring animals from space is leading to promising results \citep[e.g.][]{fretwell2014, yang2014}. Of all animal survey methods will have the lowest disturbance, but this method is still in its infancy.

In order to accurately monitor animal populations, the detections/identifications from such aerial footage must be converted into animal number densities. Achieving that requires understanding the ``completeness" of the observations, i.e. the fraction of a species in the area surveyed that each observation is sensitive to. This completeness fraction, $F_C$, will depend on environmental factors (e.g., ground temperature, vegetation coverage), species-specific factors (e.g., mean size, shape, body temperature, daily behaviour patterns) and flight-specific parameters (e.g., height above ground, angle of camera relative to ground level). All of these factors are likely to vary between different flights. Many will also vary within the same flight and even as function of pixel area within individual frames. $F_C$ is therefore a potentially complicated function of time and pixel location in a given dataset. 

The software pipeline we have developed provides a convenient way to estimate $F_C$ in environments where variations in ground temperature rather than vegetation cover is the limiting factor\footnote{See \citet{mulero2015} for a discussion on how to estimate completeness in environments with variable vegetation coverage.}. Examples of such environments would be pasture-land (such as that in which we conducted the pilot project), savannah and any generally open, tree-less plains. 

To estimate $F_C$ we replicate the technique used in astronomy to estimate how complete a given observational dataset is to stars of a given brightness. In this technique, astronomers inject fake stars with known brightness at random locations into their image and measure how many of these are successfully recovered using their detection algorithms. By repeating this process a very large number of times in a Monte-Carlo experiment and varying the brightness of the stars injected, they determine the fraction of stars they expect to recover as a function of position across the image.

To estimate $F_C$ in aerial video footage, the same outcome can be achieved by inserting fake objects of interest rather than stars into the video frames. In order to do this for the pilot project data we ran the following Monte-Carlo experiment. We first randomly selected a location in the current video frame at which to inject a fake cow or human. We then randomly selected an image cutout of a cow or human that had been previous successfully identified from a similar location in the frame in a different flight when the drone was at a similar height. We then manually added the cow or human to the image at that location and ran the detection and identification steps again to see if this cow or human was successfully identified. By repeating this process hundreds of times with different randomly selected locations and humans/cows we calculated the fraction of times humans/cows were recovered as a function of position across the image for every frame in the video footage.

We found that for the pilot data sets the completeness level varied less than 5\% across the images throughout the flight, with a few exceptions. When an object was placed close to roads the completeness level dropped to almost 0\%. This is because the roads were much hotter than the fields, so the contrast between the object and the background was much smaller (see Figure~\ref{fig:thermal_snapshots}). The completeness also dropped markedly when the object was placed too close to another object of the same type. This is the same issue in object crowding discussed above.

Our tests were conducted in the UK where the contrast between a warm animal body and the ground will be larger than in countries where day time temperatures are higher and the ground surface will be warmer as well. A previous study indicated that in South Africa detection of rhinoceros with a thermal imaging camera was best during the early morning when the contrast between the surface and the animal is highest \citep{mulero2014}. This indicates that for animal detection early morning flights might lead to the best detection and that there is therefore no need for night flights for animal detection. Night flights might, however, be needed to detect poachers. In this case exemptions from standard regulations where a drone system needs to remain visible are needed because adding LEDs to a drone for visibility would also give the location of the system away to poachers.

\subsection{Future work}
\label{sec:future}

The pilot project has demonstrated that the software pipeline has the potential to help automate the detection and identification of animals in aerial footage. The next step is to prepare the system for work in the field to begin systematic monitoring of animal populations. With help from established conservation teams, we have begun to analyse existing video footage taken in National Parks around the world. We are focusing on developing the algorithms to robustly identify megafauna in these parks, especially those which are endangered, such as rhinos. In order to build up the machine learning vector libraries for such species, we are working with Knowsley Safari Park (United Kingdom) to obtain video footage of these species from all different viewing angles as a function of time of day, vegetation coverage etc. After creating the libraries, we will begin to test them systematically on the existing footage to quantify how robustly we can detect animals, and determine the completeness levels we will be able to provide. Armed with this information we then hope to begin working with National Parks to begin a systematic monitoring programme that will automatically provide number densities of different species as a function of time and position, and alert rangers to possible poaching activities.

Although we are primarily focused on using the pipeline for conservation research, such a generic pipeline has the potential to be useful in other areas. We are therefore also actively exploring other areas in which the pipeline may be of use for scientific research or humanitarian efforts.

\section{Conclusions}
\label{sec:conclusions}

We have described a drone plus thermal-infrared camera system and software pipeline that we developed with the aim of helping monitor animal populations for conservation purposes. We demonstrated that astronomy software can be used to efficiently and reliably detect humans and animals in aerial thermal-infrared footage. Combining this astronomical detection software with existing machine learning algorithms into a single pipeline we tested the software using video footage taken in a controlled, field-like environment. We demonstrated that the pipeline works well and described how it can be used to estimate the completeness of different observational datasets for objects of a given type as a function of height, observing conditions etc. -- a crucial step in converting footage to scientifically useful information such as animal densities. We are currently taking the steps necessary to adapt the system for work in the field and hope to begin systematic monitoring of endangered species in the near future.

\section{Acknowledgements}
\label{sec:acknowledgements}

We thank Trevor Smith, the owner of Arrowe Brook Farm, for letting us fly the drone system over his fields. The data obtained from those flights form the basis of the pilot project described in $\S$~\ref{sec:pilot}. SNL acknowledges continued support from the Astrophysics Research Institute who purchased the initial drone plus camera system. SNL also acknowledges support from Liverpool John Moores University via a Research Spotlight Award and studentship funding. JHK acknowledges financial support from the Spanish Ministry of Economy and Competitiveness (MINECO) under grant number AYA2013-41243-P.  SNL would like to thank Prof. Nate Bastian for supplying test drones at the onset of the project which were used for flight practice. 

\bibliography{longmore_ijrs_2016}
\end{document}